\documentclass[aps,notitlepage,nofootinbib,twocolumn,superscriptaddress,10pt]{revtex4-1}

\newcommand{\bA}{{\bf A}}
\newcommand{\bB}{{\bf B}}
\newcommand{\bE}{{\bf E}}

\newcommand{\bJ}{{\bf J}}
\newcommand{\sigmahat}{\hat{\sigma}}
\newcommand{\kappahat}{\hat{\kappa}}
\newcommand{\alphahat}{\hat{\alpha}}
\newcommand{\Shat}{\hat{S}}
\newcommand{\kb}{k_\textrm{B}}

\newcommand{\beq}{\begin{eqnarray}}
\newcommand{\eeq}{\end{eqnarray}}
\newcommand{\beqq}{\begin{eqnarray*}}
\newcommand{\eeqq}{\end{eqnarray*}}

\newcommand{\be}{\begin{equation}}
\newcommand{\ee}{\end{equation}}
\newcommand{\barr}{\begin{array}}
\newcommand{\earr}{\end{array}}

\newcommand{\sign}{\text{sign}}

\newcommand{\ve}{{\varepsilon}}

\usepackage[colorlinks=true,citecolor=blue,linkcolor=blue]{hyperref}
\usepackage{graphicx}
\usepackage{dcolumn}
\usepackage{bm}
\usepackage{color}
\usepackage{tabularx}
\usepackage{comment}

\usepackage{amsmath}

\begin{document}

\begin{titlepage}

\widetext

\title{Thermoelectric Hall conductivity and figure of merit in Dirac/Weyl materials}

\author{Vladyslav Kozii}
\thanks{These two authors contributed equally.}
\author{Brian Skinner}
\thanks{These two authors contributed equally.}
\author{Liang Fu}
\affiliation{Department of Physics, Massachusetts Institute of Technology, Cambridge,
Massachusetts 02139, USA}

\setcounter{equation}{0}
\setcounter{figure}{0}
\setcounter{table}{0}

\makeatletter
\renewcommand{\theequation}{S\arabic{equation}}
\renewcommand{\thefigure}{S\arabic{figure}}
\renewcommand{\thetable}{S\Roman{table}}
\renewcommand{\bibnumfmt}[1]{[S#1]}
\renewcommand{\citenumfont}[1]{S#1}

\date{\today}

\begin{abstract}
We calculate the thermoelectric response coefficients of three-dimensional Dirac or Weyl semimetals as a function of magnetic field, temperature, and Fermi energy.  We focus in particular on the thermoelectric Hall coefficient $\alpha_{xy}$ and the Seebeck coefficient $S_{xx}$, which are well-defined even in the dissipationless limit.  We contrast the behaviors of $\alpha_{xy}$ and $S_{xx}$ with those of traditional Schr\"{o}dinger particle systems, such as doped semiconductors. Strikingly, we find that for Dirac materials $\alpha_{xy}$ acquires a constant, quantized value at sufficiently large magnetic field, which is independent of the magnetic field or the Fermi energy, and this leads to unprecedented growth in the thermopower and the thermoelectric figure of merit. We further show that even relatively small fields, such that $\omega_c \tau \sim 1$ (where $\omega_c$ is the cyclotron frequency and $\tau$ is the scattering time), are sufficient to produce a more than $100\%$ increase in the figure of merit.

\end{abstract}


\maketitle

\draft

\vspace{2mm}

\end{titlepage}

\section{Introduction}

In an electrically conductive system at finite temperature, the quasiparticle excitations that carry electric current also carry heat current.  The magnitude of the heat current density $\bJ^Q$ at a particular value of the electric field is described by the Peltier conductivity tensor $\alphahat$.  In particular, in the presence of an electric field $\bE$ and a gradient of temperature $T$, the electric and thermal current densities are given by \cite{AMbook}
\beq
\bJ & = & \sigmahat \bE - \alphahat \nabla T \\
\bJ^Q & = & T \alphahat \bE - \kappahat \nabla T.
\eeq
Here, $\bJ$ is the electric current density, $\sigmahat$ is the electrical conductivity tensor, and $\kappahat$ is the thermal conductivity tensor.  The Peltier conductivity tensor $\alphahat$ is related to the thermoelectric tensor $\hat{S}$ by $\hat{S} = \sigmahat^{-1} \alphahat$.

At temperatures much lower than the Fermi temperature, the thermoelectric response coefficients $\alphahat$ and $\hat{S}$ due to charge carriers are typically proportional to $k_B T/E_F \ll 1$, where $k_B$ is the Boltzmann constant and $E_F$ is the Fermi energy~\cite{AMbook}.  $E_F$ is typically very large in a good metal, which leads to a small magnitude of the thermoelectric response.  Thus the thermoelectric response coefficients are typically appreciable only in systems with relatively low Fermi energy, for example in doped semiconductors.

During the last decade there has been a surge of interest in the thermoelectric properties of materials with topological or otherwise unconventional band structure. (See, for example, Refs.\ \onlinecite{ Kimgraphene, Behniabook, Shigraphene, Checkelskygraphene, XiaoNiu06, FauqueBi2Se3, Potternu12, OngPbSnSe}.) The electronic contribution to the thermoelectric response coefficients $\alphahat$ and $\hat{S}$ reflect the properties of the quasiparticle dispersion.  In this way, measuring $\alphahat$ or $\hat{S}$ provides a way of studying the nature of electronic quasiparticles.

\begin{figure}[htb]
\centering
\includegraphics[width=.9 \columnwidth]{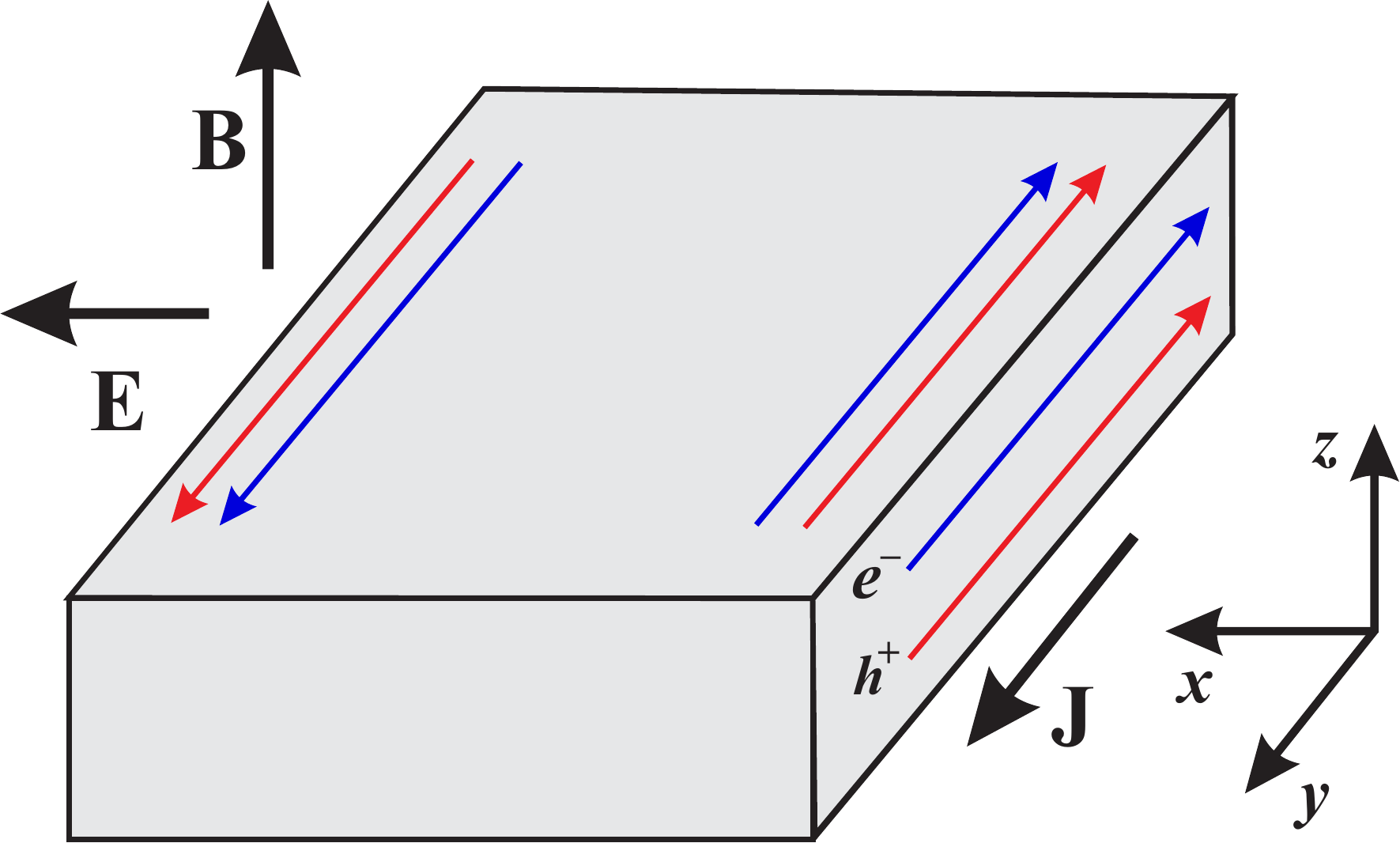}
\caption{Schematic illustration of electron ($e^{-}$) and hole ($h^{+}$) currents along edge states in the dissipationless limit, where the electric field $\bE$ is perpendicular to the electric current $\bJ$ and to the magnetic field $\bB$.}
\label{fig:edgestates}
\end{figure}

Experiments on transverse thermoelectric response commonly focus on the
Nernst effect, in which a voltage gradient is measured in the direction \emph{transverse} to an applied temperature gradient (e.g.,\ Refs.\ \cite{HeremansNernst, OngNernst, BehniaNernst}).  However, in a sufficiently strong magnetic field even the diagonal component of the thermopower (the Seebeck coefficient $S_{xx}$) can take a value that is independent of the disorder scattering.  In fact, in a recent paper, we showed that in three-dimensional Dirac or Weyl semimetals this large-field value of $S_{xx}$ can be enormously enhanced by a sufficiently strong magnetic field. \cite{SkinnerFu}

The usefulness of the Nernst coefficient $S_{xy}$ for studying the intrinsic band structure, and the independence of the Seebeck coefficient $S_{xx}$ on disorder at large field, can both be viewed as a consequence of the off-diagonal component of $\alphahat$ having a large dissipationless contribution.  In this paper, we study this off-diagonal component $\alpha_{xy}$, which we refer to as the ``thermoelectric Hall coefficient'', in detail.  We calculate its value for three-dimensional Dirac/Weyl semimetals as a function of magnetic field, temperature, and carrier density, and we contrast the results with the behavior of $\alpha_{xy}$ for conventional Schr\"{o}dinger quasiparticles (studied in detail in Ref.~\cite{Oganesyan10}), for which the kinetic energy varies quadratically with momentum.  In both cases, the value of $\alpha_{xy}$ attains a maximum at a particular value of magnetic field.  Strikingly, however, for Dirac/Weyl semimetals the value of $\alpha_{xy}$ settles into a plateau at large magnetic field, such that the quantity $\alpha_{xy} v_F/T$ is quantized, where $v_F$ is the Fermi velocity in the field direction.  This is shown in Fig.~\ref{fig:alphaxyD}.

In the remainder of this paper, we calculate $\alpha_{xy}$ using the relation
\be 
\alpha_{xy} = \frac{J_y^Q}{T E_x},
\label{eq:alphaxydef}
\ee
in which the temperature is taken to be uniform across the system and the electric field $\bE$ is taken to be in the $x$ direction.  We calculate the thermoelectric Hall response using two complementary approaches. First, we consider the dissipationless limit, where the transport scattering time diverges and all heat current is provided by quantum Hall edge channels (see Fig.\ \ref{fig:edgestates}).  Second, we use a quasiclassical Boltzmann equation description to consider the case where the transport scattering time $\tau$ is finite.  These two descriptions agree in the case where $\omega_c \tau \gg 1$, where $\omega_c$ is the cyclotron frequency, provided that multiple Landau levels are occupied.
Finally, we also use the Boltzmann equation to study the Seebeck coefficient $S_{xx}$. While $S_{xx}$ in the dissipationless limit, which corresponds to high fields $\omega_c \tau \gg 1$, was exhaustively studied in Ref.~\onlinecite{SkinnerFu}, here we focus on the case of small fields $\omega_c \tau \sim 1$. We show that even relatively low fields are sufficient to enhance $S_{xx}$ in Dirac materials, increasing the figure of merit of thermoelectric devices by $\approx 100\%$. This result is in contrast to the case of Schr\"{o}dinger materials, where $S_{xx}$ remains constant at small fields if one assumes an energy-independent value of $\tau$. 
We focus everywhere in this paper on the ``electron diffusion'' contribution to the thermopower; the effects of phonon drag are left for a future work.

The remainder of the paper is organized as follows. Section \ref{sec:dissipationless} gives a general expression for $\alpha_{xy}$ in the dissipationless limit, which largely recapitulates the canonical derivations in Refs.\ \onlinecite{Halperin82, GirvinJonson, Oganesyan10}. Section \ref{sec:QC} discusses the quasiclassical approximation, and gives a general expression for $\alpha_{xy}$ in terms of the Hall conductivity, which we describe using the Boltzmann equation.  Section \ref{sec:Sch} describes the results for Schr\"{o}dinger particles, using both approximations, and Sec.\ \ref{sec:Dirac} gives the results for Dirac quasiparticles.  We close in Sec.\ \ref{sec:discussion} with a summary and discussion.

\begin{figure}[htb]
\centering
\includegraphics[width=.8 \columnwidth]{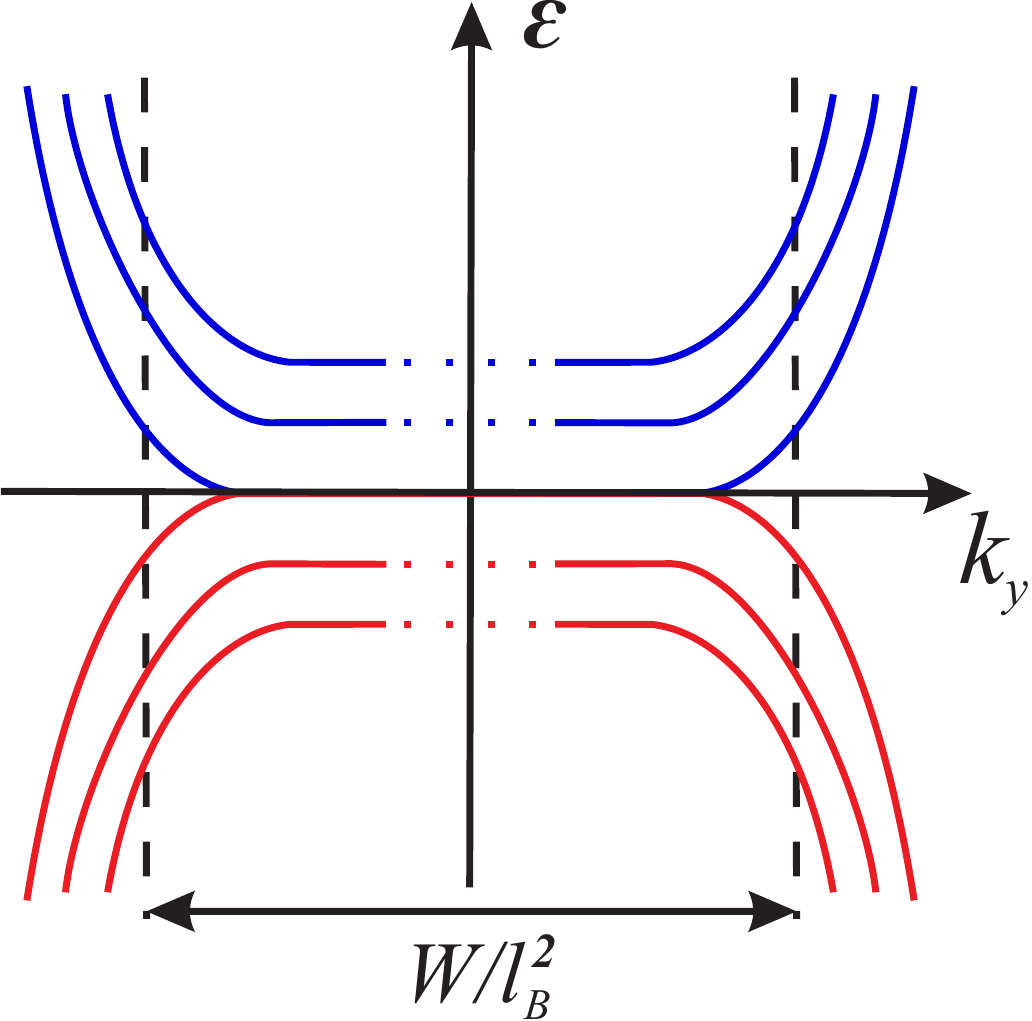}
\caption{Landau levels $\ve_n(k_y,k_z)$ in the presence of a confining potential in the $x$-direction. In the Schr\"{o}dinger case the states with negative energies $\ve<0$ (red lines) are absent. Dashed lines denote momenta corresponding to states located near the boundaries of the Hall brick, $k_y = \pm W/2l_B^2$. We assume that the magnetic length is much smaller than the width of the brick, $l_B \ll W$, such that bulk bands remain nearly flat. }
\label{fig:Landaulevels}
\end{figure}

\section{Dissipationless limit}
\label{sec:dissipationless}

In cases when the scattering rate is small compared to the cyclotron frequency, $\omega_c \tau \gg 1$, both the Hall conductivity $\sigma_{xy}$ and the thermoelectric Hall coefficient $\alpha_{xy}$ can be calculated using the quantum Hall edge formalism developed by Halperin~\cite{Halperin82} and by Girvin and Jonson~\cite{GirvinJonson}. For simplicity, we focus here on the ``Hall brick'' geometry (see Fig.~\ref{fig:edgestates}), in which the sample is taken to have a finite extent $W$ in $x$-direction. The magnetic field is taken to be along the $z$-direction.  We describe the electron eigenstates using the Landau gauge $\bA = x B \hat y$, where $\bA$ is the vector potential, so that the states are parameterized by their quasimomenta $k_y$ and $k_z$. The corresponding eigenfunctions are centered at a lateral position $x_0(k_y) = k_y l_B^2$, where $l_B = (\hbar/eB)^{1/2}$ is the magnetic length. 

In the absence of a confining potential in the $x$-direction, the energy levels are highly degenerate and do not depend on $k_y$.  The corresponding electron energy is then given by $\ve = \ve_n^0(k_z)$, where $n$ is the Landau level index. The function $\ve_n^0(k_z)$ depends on the quasiparticle dispersion, as we describe below for the cases of Schr\"odinger and Dirac particles. In the presence of a confining potential in $x$-direction, however, the energy levels disperse with $k_y$ also, so that $\ve = \ve_n(k_y,k_z)$, as illustrated in Fig.~\ref{fig:Landaulevels}. 

The total current in the $y$-direction is given by 
\be
I_y = \frac{e}{L_y} \sum_{\text{all states}} v_y n_F(\ve - \mu),
\ee
where $L_y$ is the size of the brick in the $y$-direction, $v_y$ is the $y$-component of the velocity of a state with energy $\ve$, $n_F(\ve) = [1 + \exp(\ve/ k_B T)]^{-1}$ is the Fermi-Dirac distribution, and $\mu$ is the electrochemical potential. To derive an explicit expression for the current, we recall that the electron velocity in $y$-direction is given simply by $v_y = (1/\hbar) \partial \ve(k_y , k_z)/ \partial k_y$.  The presence of an electrostatic potential difference $V_x$ between the two edges of the brick implies a spatial variation of the electrochemical potential $\mu$. Given that the states with different $k_y$ are centered at different positions $x_0(k_y) = k_y l_B^2,$ this spatial variation can be cast into the effective dependence of $\mu$ on $k_y$, i.e., $\mu(k_y) \simeq \mu_0 + e V_x x/W = \mu_0 + e V_x l_B^2 k_y/W$. Here $\mu_0$ is the electrochemical potential in the absence of an electric field. Expanding then the Fermi distribution to the first order in $V_x$, we find
\begin{multline}
I_y = -\frac{e^2}{\hbar} \frac{V_x l_B^2}{W L_y} \sum_{k_z, k_y,n} N_n k_y \frac{\partial \ve_n(k_y, k_z)}{\partial{k_y}} \times  \\ \times  \frac{\partial }{\partial \ve} n_F[\ve_n(k_y, k_z) - \mu_0], \label{Eq:Iy1}
\end{multline}
where $N_n$ is the degeneracy of the level with energy $\ve_n$ (for a given $k_y$ and $k_z$), and for brevity we will suppress the subscript $0$ in $\mu_0$ hereafter.


If the magnetic field is sufficiently strong that $W\gg l_B$, the energy bands in the bulk remain nearly flat as a function of $k_y$ (up to exponentially small corrections), and the corresponding contribution to the total current $I_y$ can be neglected due to the smallness of the velocity $v_y$. Consequently, the most significant contribution to $I_y$ is due to the familiar quantum Hall edge states, and one can set $k_y\approx \pm W/2l_B^2$ in Eq.~(\ref{Eq:Iy1}). This assumption allows us to change the summation variable $k_y$ to $\ve$. Performing then the integration over $\ve$ explicitly, we find
\be
\sigma_{xy} = \frac{I_y}{V_x L_z} = \sigma_{xy}^e - \sigma_{xy}^h, \label{Eq:sigma1}
\ee
where $L_z$ is the size of the brick in the $z$ direction and the electron and hole contributions to the conductivity, $\sigma_{xy}^e$ and $\sigma_{xy}^h$, respectively, are given by
\begin{align}
&\sigma_{xy}^e = \frac{e^2}{2\pi \hbar} \int_{-\infty}^{\infty}\frac{d k_z}{2\pi}\sum_{n:\ve_n^0>0} N_n \cdot n_F\left[ \ve_n^0(k_z) -\mu \right], \label{Eq:sigma} \\ &\sigma_{xy}^h = \frac{e^2}{2\pi \hbar} \int_{-\infty}^{\infty}\frac{d k_z}{2\pi}\sum_{n:\ve_n^0<0} N_n \cdot \left(1-n_F\left[ \ve_n^0(k_z) -\mu \right]\right). \nonumber
\end{align}
Strictly speaking, the bulk value of the Landau level energy $\ve_n^0(k_z)$ in the above expression should be substituted with $\ve_n^0(k_y=0,k_z)$; however, in the limit $W\gg l_B$ considered in this paper, they are approximately equal, $\ve_n^0(k_y=0,k_z) \approx \ve_n^0(k_z)$. The second contribution in Eq.~(\ref{Eq:sigma}), $\sigma_{xy}^h$, represents a sum over negative-energy Landau levels in the valence band.  For Schr\"odinger particles, where the valence band is very far from the chemical potential, the contribution $\sigma_{xy}^h$ can be neglected.  However, the contribution from these negative Landau levels plays a significant role for Dirac/Weyl semimetals at finite temperature and sufficiently large magnetic field, as we show below.

In order to describe the Hall conductivity $\sigma_{xy}$ at a given magnetic field and electron concentration $n_0$, one should introduce the self-consistency condition for the chemical potential $\mu$:
\be
\int_0^{\infty}d\ve \, \nu(\ve) n_F(\ve - \mu) - \int_{-\infty}^{0}d\ve \,\nu(\ve) \left[ 1-n_F(\ve - \mu)  \right] = n_0. \label{Eq:mu}
\ee
Here, the first term on the left-hand side represents the number of electrons per unit volume, and the second term is the number of holes.  The bulk density of states $\nu(\ve)$ is given by 
\be
\nu(\ve) = \frac{e B}{2\pi \hbar}\sum_{k_z, n} N_n \cdot \delta \left[\ve - \ve_n^0(k_z)\right], \label{Eq:DOS}
\ee
where $e B / 2\pi \hbar$ is the number of flux quanta per unit area.  The second term in Eq.~(\ref{Eq:mu}) is absent for Schr\"odinger particles, since $\nu(\ve<0) =0$ in that case.

Combining together Eqs.~(\ref{Eq:sigma1})--(\ref{Eq:DOS}), one easily finds the famous result for the Hall conductivity,
\be
\sigma_{xy} = \frac{e n_0}{B},
\ee
which is typically explained  classically by noting that in the dissipationless limit the electron current is entirely due to the transverse $\bE \times \bB$ drift of all electrons with the drift velocity $v_d = E_x/B$ in the $y$-direction. 

Analogously, one can derive a general expression for the thermoelectric Hall coefficient $\alpha_{xy}$. In the presence of a potential difference $V_x$, the total heat current in $y$-direction is equal to
\begin{multline}
I_y^Q = -\frac{e}{\hbar} \frac{V_x l_B^2}{ W L_y} \sum_{k_z, k_y,n} N_n k_y \times [\ve_n(k_y, k_z) - \mu] \\ \times \frac{\partial \ve_n(k_y, k_z)}{\partial{k_y}}  \frac{\partial }{\partial \ve} n_F[\ve_n(k_y, k_z) - \mu].
\end{multline}
This equation differs from Eq.\ (\ref{Eq:Iy1}) by the factor $\ve_n(k_y, k_z) - \mu$ within the sum, which describes the energy carried by each electron or hole state.
Assuming, as with $I_y$, that the main contribution to the heat current is due to the edges at $k_y\approx \pm W/2l_B^2,$ one can easily perform integration over $k_y$, resulting in 
\be
\alpha_{xy}(B,T) = \frac{I_y^Q}{T V_x L_y} = \frac{e}{2\pi \hbar L_z} \sum_{n,k_z} N_n s\left( \frac{\ve_n^0(k_z) - \mu}{k_B T} \right). \label{Eq:alpha}
\ee
Here we have introduced the entropy per electron state 
\begin{align}
s(x) &= -k_B \left[ n_F \ln n_F + (1-n_F)\ln (1-n_F)  \right] = \nonumber \\ &= k_B \left[ \ln\left(1+ e^x  \right) - \frac{x}{1+e^{-x}} \right]. \label{Eq:entropy}
\end{align}
This connection between $\alpha_{xy}$ and entropy has previously been discussed for Schr\"{o}dinger particles~\cite{Oganesyan10}, and here we demonstrate that it is also valid more generically, and can be applied, for example, to the case of Dirac particles.  

Finally, we note that the Seebeck coefficient $S_{xx}$, which plays a crucial role in determining the figure of merit of thermoelectric devices~\cite{SkinnerFu}, is generally defined as 

\be
S_{xx} =  S_{yy} = \frac{I_y^Q}{T I_y} = \frac{\alpha_{xx}\sigma_{xx} + \alpha_{xy} \sigma_{xy}}{\sigma_{xx}^2 + \sigma_{xy}^2}. \label{Eq:Sxx}
\ee
In the dissipationless limit, where $\sigma_{xy} \gg \sigma_{xx}$, it has the simple form~\cite{SkinnerFu}
\be
S_{xx} =  \frac{\alpha_{xy}}{\sigma_{xy}} =  \frac{B}{2\pi\hbar n_0} \sum_{n,k_z} N_n s\left( \frac{\ve_n^0(k_z) - \mu}{k_B T} \right). 
\ee


\section{Quasiclassical approximation}
\label{sec:QC}
 
The approach used in the previous section is universal in the strong magnetic field limit, $\omega_c \tau \gg 1$. However, at small magnetic field, this condition is violated, and quasiparticle scattering must be taken into account. The most straightforward way to account for the finite scattering rate is within the Boltzmann quasiclassical theory. In this description the general expressions for the conductivity and the thermoelectric coefficients (both longitudinal and Hall parts) are
\begin{align}
\sigma_{xx(xy)} &=  \int d\ve \left( -\frac{\partial n_F}{\partial \ve} \right) \sigma_{xx(xy)}(\ve)\nonumber \\ \alpha_{xx(xy)} &= \frac{1}{eT} \int d\ve \, (\ve - \mu) \left( -\frac{\partial n_F}{\partial \ve} \right) \sigma_{xx(xy)}(\ve). 
\label{Eq:QC}
\end{align}
Within the Boltzmann approach, the energy-dependent conductivity is given by
\be
\left( \begin{array}{c} \sigma_{xx}(\ve) \\  \sigma_{xy}(\ve) \end{array} \right)= \frac{1}{3} \frac{ e^2 \nu(\ve) v_F^2(\ve)  \tau(\ve) }{1+\omega_c^2(\ve) \tau^2(\ve)} \left( \begin{array}{c} 1 \\  \omega_c(\ve) \tau(\ve)    \end{array} \right).
\ee
In should be emphasized that, in general, the Fermi velocity $v_F$, the cyclotron frequency $\omega_c$, and the scattering time $\tau$ (in addition to the density of states $\nu$) are functions of energy, and they depend on the type of particle dispersion and on the mechanism for quasiparticle scattering. In what follows, however, we focus for simplicity on a model with constant (energy-independent) scattering time $\tau$. 

In the limit when both the cyclotron energy and the temperature are smaller than the Fermi energy, $\hbar \omega_c, k_B T \ll E_F,$ one can evaluate the integrals in Eqs.~(\ref{Eq:QC}) using a Sommerfeld expansion, yielding
\begin{align}
\sigma_{xx(xy)} &\approx \left.  \sigma_{xx(xy)}(\ve) \right\vert_{\ve = E_F }, \label{Eq:Sommerfeld} \\ \alpha_{xx(xy)} & \approx  \left. \frac{\pi^2}3 \frac{k_B^2 T}e \frac{d}{d \ve}  \sigma_{xx(xy)}(\ve) \right\vert_{\ve = E_F }. \nonumber 
\end{align}
The Seebeck coefficient can then be found by inserting these equations into Eq.~(\ref{Eq:Sxx}).  In the limit $\hbar \omega_c, k_B T \ll E_F$, the result can be written as
\begin{align}
S_{xx} &\approx \left. \frac{\pi^2}6 \frac{k_B^2 T}{e} \frac{d}{d \ve}  \ln\left( \sigma_{xx}^2(\ve) + \sigma_{xy}^2(\ve)  \right) \right\vert_{\ve = E_F } \nonumber \\ &= \left. \frac{\pi^2}6 \frac{k_B^2 T}{e} \frac{d}{d \ve}  \ln\left( \frac{\nu^2(\ve) v_F^4(\ve) \tau^2(\ve)}{1 + \omega_c^2(\ve) \tau^2(\ve)} \right) \right\vert_{\ve = E_F } \label{Eq:SxxQC}
\end{align}
The first equation is equivalent to the longitudinal component of the usual Mott formula for the thermopower at low temperature,
\be 
\Shat = \frac{\pi^2}{3} \frac{k_B^2 T}{e} \sigmahat^{-1} \left. \frac{d \sigmahat}{d \ve} \right|_{\ve = E_F}.
\nonumber 
\ee

The quasiclassical expressions (\ref{Eq:QC})--(\ref{Eq:SxxQC}) are applicable when a large number of Landau levels is filled, i.e., at sufficiently weak magnetic fields that $\hbar \omega_c \ll E_F$. However, if the scattering time $\tau$ is sufficiently long, then there exists a window of magnetic fields such that $1/\tau \ll \omega_c \ll E_F/\hbar$.  The first inequality in this chain implies that transport is essentially dissipationless, while the second implies that the quasiclassical approach is valid.  Thus, in this window of magnetic fields, the quasiclassical result coincides with the dissipationless result from Sec.\ \ref{sec:dissipationless}.  By merging the two descriptions we can therefore obtain the result for $\alpha_{xy}$ and $S_{xx}$ over the whole range of magnetic field.

\section{Schr\"{o}dinger particles}
\label{sec:Sch}

We now apply the general formalism from the previous two sections to the familiar case of Schrodinger particles, as considered, e.g., in Ref.~\onlinecite{Oganesyan10}. This scenario is realized, for example, in heavily doped semiconductors. Assuming, for simplicity, an isotropic band with mass $m$, the bulk Landau levels have energy $\ve_n^0(k_z)$ given by
\be
\ve_n^0(k_z) = \hbar \omega_c \left(n+\frac12\right) + \frac{\hbar^2 k_z^2}{2m},
\ee
where $n$ is a non-negative integer and the cyclotron frequency $\omega_c = e B /m$. Here we also neglect the effects of Zeeman splitting, which amounts to an assumption that the effective $g$-factor is small. In this case, the degeneracies of all Landau levels (at fixed $k_y$) are the same, and are given simply by the number $N_f$ of electron flavors (which includes the spin degeneracy). The density of states is then given by
\be
\nu^S(\ve) = \frac{N_f B e \sqrt{2m}}{(2\pi\hbar)^2}\text{Re} \sum_{n=0}^{\infty} \frac1{\sqrt{\ve - \hbar \omega_c (n+1/2)}},
\ee
where the superscript $S$ stands for ``Schr\"{o}dinger''. 

\begin{figure}[htb!]
\centering
\includegraphics[width=1.0 \columnwidth]{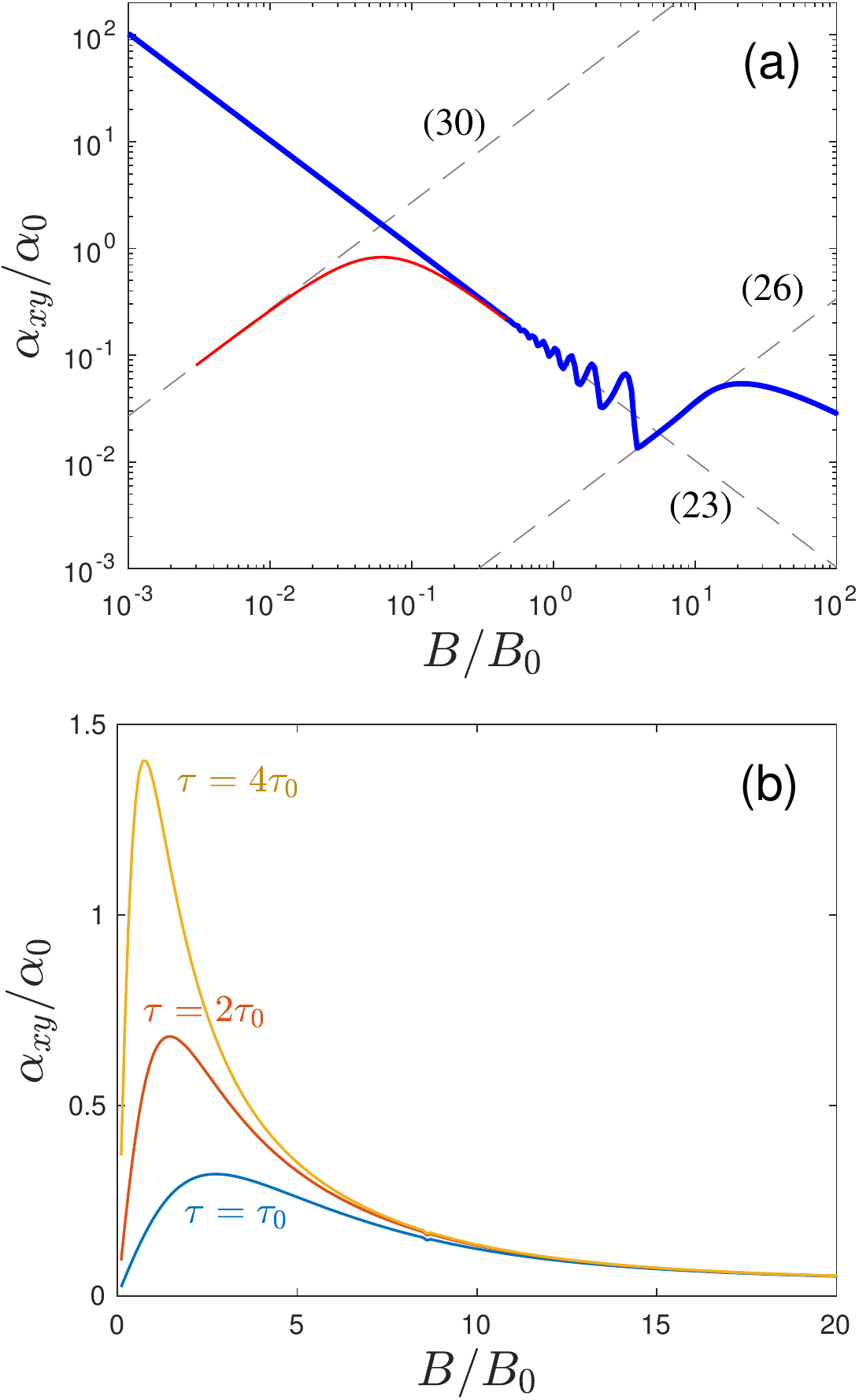}
\caption{The thermoelectric Hall coefficient $\alpha_{xy}$ of Schr\"{o}dinger particles in three dimensions as a function of magnetic field $B$.  (a) A double-logarithmic plot of $\alpha_{xy}$, showing both the result in the dissipationless limit (thick blue line) and the semiclassical result (thin red line) corresponding to a scattering time $\tau = 50 (v_F n_0^{1/3})^{-1}$, with $v_F = \hbar (6 \pi^2 n_0/N_f)^{1/3}/m$ being the Fermi velocity at zero magnetic field. The labeled dashed lines show the limiting results of Eqs.\ (\ref{Eq:alphaSclasslowB}), (\ref{Eq:alphaS1}), and (\ref{Eq:alphaS2}), respectively.  The temperature is taken to be $T = 0.1 \hbar v_F n_0^{1/3}/k_\textrm{B}$.  (b) $\alpha_{xy}$ vs $B$ in linear scale, as given by the semiclassical calculation, calculated for a large enough temperature that quantum oscillations are washed out ($T = 1 \times \hbar v_F n_0^{1/3}/k_\textrm{B}$).  Different curves are labeled according to their value of the scattering time $\tau$, with $\tau_0 = (v_F n_0^{1/3})^{-1}$.   In both plots, the units of magnetic field are $B_0 = \hbar n_0^{2/3}/e$, and units of $\alpha_{xy}$ are $\alpha_0 = e k_\textrm{B} n_0^{1/3}/\hbar$. 
}
\label{fig:alphaxyS}
\end{figure}

Using the general expression~(\ref{Eq:alpha}) for the dissipationless limit, we find for the thermoelectric Hall coefficient $\alpha_{xy}$
\be
\alpha_{xy}^S =  \frac{e N_f}{2\pi \hbar} \sum_{n=0}^{\infty} \int_0^{\infty} \frac{d k_z}{\pi} s\left( \frac{\ve_n^0(k_z) - \mu}{k_B T} \right), \label{Eq:alphaS}
\ee
where the function $s(x)$ is defined by Eq.~(\ref{Eq:entropy}), and the chemical potential $\mu$ as a function of density $n_0$, temperature $T$, and magnetic field $B$ must be self-consistently determined from Eq.~(\ref{Eq:mu}). The behavior of $\alpha_{xy}^S$ as a function of magnetic field is shown in Fig.~\ref{fig:alphaxyS}.

Limiting cases of the general expression (\ref{Eq:alphaS}) for the dissipationless limit can be understood as follows. For definiteness, we focus on the case when the temperature is much smaller than the Fermi energy, $k_B T\ll E_F = \left( 3\pi^2 \hbar^3 n_0/N_f m \sqrt{2m} \right)^{2/3}$. At sufficiently small magnetic fields that $\hbar \omega_c \ll E_F$, the density of states remains unchanged to the leading order in magnetic field, $\nu(\ve \gg \hbar \omega_c) \approx N_f m \sqrt{2 m \ve}/2\pi^2 \hbar^3,$ and the chemical potential coincides with the Fermi energy, $\mu \approx E_F$. In this limit we find for the thermoelectric Hall coefficient
\be
\alpha_{xy}^{S}  \approx \left( \frac{N_f \pi}6  \right)^{2/3}\frac{n_0^{1/3} m k_B^2 T}{\hbar^2 B}. \label{Eq:alphaS1}
\ee

On the other hand, when the magnetic field is large enough that $\hbar \omega_c$ becomes larger than the Fermi energy, only the states within the zeroth Landau level contribute to transport. In this case, the density of states associated with the lowest Landau level is
\be
\nu = \frac{N_f Be\sqrt{2m}}{(2\pi\hbar)^2} \cdot \frac1{\sqrt{\ve - \hbar \omega_c/2}}
\ee
and the chemical potential is given by
\be
\mu - \frac{\hbar \omega_c}2  \approx \frac{2 \pi^4 \hbar^4 n_0^2}{m e^2 B^2 N_f^2} \ll \hbar \omega_c. 
\ee
In the limit of small temperatures $k_B T \ll \mu - \hbar \omega_c/2$, the entropy $s \approx (\pi^2/3) k_B^2 T \nu(\mu)$, so that
the thermoelectric Hall coefficient is
\be
\alpha_{xy}^S \approx \frac{e^2 k_B^2 T N_f^2 m B}{12 \pi^2 \hbar^4 n_0}. \label{Eq:alphaS2}
\ee
This result is valid when the magnetic field is in the range $\hbar n_0^{2/3}/e \ll B \ll \hbar^2 n_0 / e \sqrt{m k_B T}.$ 

When the magnetic field is increased even further, so that $B \gg \hbar^2 n_0 / e \sqrt{m k_B T}$, the Fermi energy relative to the bottom of the lowest Landau level becomes smaller than $k_B T$.  In this limit the chemical potential becomes negative (as in a classical ideal gas) with respect to the bottom of the lowest band:
\be
\mu - \frac{\hbar \omega_c}2  \approx k_B T  \ln \left[ \frac{(2\pi\hbar)^2 n_0}{B e N_f \sqrt{2\pi m k_B T}} \right] < 0.
\ee
In this limit electrons are well described by a classical Boltzmann distribution, leading to 
\be
\alpha_{xy}^S  \approx \frac{n_0 k_B}{B} \ln \left[ \frac{N_f B e \sqrt{2\pi m k_B T}}{(2\pi\hbar)^2 n_0} \right]. \label{Eq:alphaS3}
\ee

Equations (\ref{Eq:alphaS})--(\ref{Eq:alphaS3}) are valid when electron scattering can be completely ignored. Equation (\ref{Eq:alphaS1}), in particular, implies that in this dissipationless limit the value of $\alpha_{xy}^S$ diverges as $1/B$ in the limit of small magnetic field.  In reality, however, this divergence of $\alpha_{xy}^S$ is cut off by the finite scattering time, which truncates the divergence when the magnetic field is small enough that $\omega_c \tau < 1$. This truncation can be described using the quasiclassical approach developed in Sec.~\ref{sec:QC}. The result at low temperatures $k_B T \ll E_F$ can be obtained directly from the Sommerfeld expansion, Eq.~(\ref{Eq:Sommerfeld}), with the Fermi velocity given by $v_F(\ve) = \sqrt{2\ve/m}$ and the density of states given by its zero-field value $\nu(\ve) \approx N_f m \sqrt{2 m \ve}/2\pi^2 \hbar^3$.  If one assumes an energy-independent scattering time $\tau$, then $\alpha_{xy}$ is given by
\be
\alpha_{xy}^{SQC}  \approx \left( \frac{N_f \pi}6  \right)^{2/3}\frac{n_0^{1/3}  k_B^2 T e}{\hbar^2} \frac{\omega_c \tau^2}{1 + \omega_c^2 \tau^2}. \label{Eq:alphaSclass}
\ee

As expected, Eq.~(\ref{Eq:alphaSclass}) reproduces the dissipationless result of Eq.~(\ref{Eq:alphaS1}) in the limit of weak disorder, $\omega_c \tau \gg 1$. At lower magnetic fields, it smoothly crosses over to
\be 
\alpha_{xy}^{SQC}  \approx \left( \frac{N_f \pi}6  \right)^{2/3}\frac{n_0^{1/3}  k_B^2 T e^2 \tau^2 B}{\hbar^2 m} . \label{Eq:alphaSclasslowB}
\ee
Thus, the value of $\alpha_{xy}$ attains a maximum at $\omega_c \tau = 1$, as can be seen in Fig.\ \ref{fig:alphaxyS}.

\section{Dirac particles}
\label{sec:Dirac}

In this section, we discuss in detail the thermoelectric Hall coefficient for three-dimensional Dirac materials, which have an energy-independent Fermi velocity $v_F$ and are the main focus of this paper. If one assumes, for simplicity, that $v_F$ is isotropic, then the Landau levels in the bulk are described by \cite{JeonDiracLLs}
\be
\ve_n^0(k_z) = \sign(n) \cdot v_F\sqrt{2 e \hbar B |n| + \hbar^2 k_z^2},
\ee
where $n$ is an integer (positive or negative).
All levels with $n \ne 0$ (and fixed $k_y$ and $k_z$) have the same degeneracy $N_f$, which is equal to the number of Weyl nodes in Weyl semimetals and is equal to twice the number of nodes in Dirac semimetals. It should be noted that $N_f$ is always even because of the fermion doubling theorem. The level with $n=0$, however, requires extra care. At non-zero $k_z$,  the $n = 0$ Landau level splits into two levels $\ve_{\pm}^0(k_z) = \pm v_F \hbar |k_z|$, each of which has degeneracy $N_f/2$. With this precaution, the density of states is given by
\be
\nu^D(\ve) = \frac{N_f B e}{2\pi^2 \hbar^2 v_F} \left(\frac12 + \text{Re} \sum_{n=1}^{\infty} \frac{|\ve|}{\sqrt{\ve^2 - 2 \hbar v_F^2 e B n}}  \right),
\ee
where the index $D$ stands for ``Dirac''. 

\begin{figure}[htb!]
\centering
\includegraphics[width=1.0 \columnwidth]{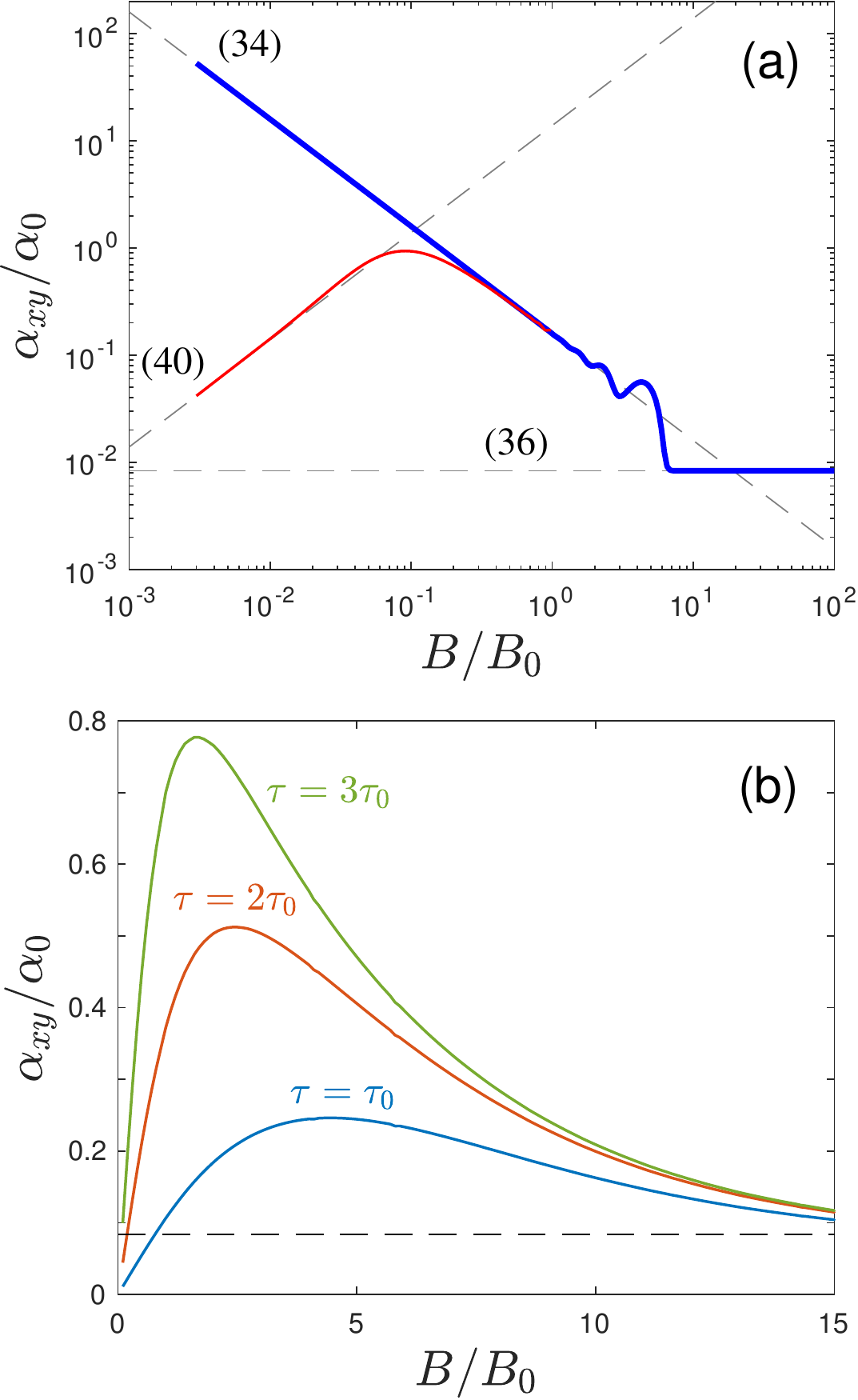}
\caption{The thermoelectric Hall coefficient $\alpha_{xy}$ of a three-dimensional Dirac/Weyl semimetal as a function of magnetic field $B$.  (a) A double-logarithmic plot of $\alpha_{xy}$, showing both the result in the dissipationless limit (thick blue line) and the semiclassical result (thin red line) corresponding to a scattering time $\tau = 50 (v_F n_0^{1/3})^{-1}$.  The labeled dashed lines show the limiting results of Eqs.\ (\ref{eq:alphaxyDsmallB}), (\ref{Eq:alphaD1}), and (\ref{Eq:alphaD2}), respectively.  The temperature is taken to be $T = 0.1 \hbar v_F n_0^{1/3}/k_\textrm{B}$.  (b) $\alpha_{xy}$ vs $B$ in linear scale, as given by the semiclassical calculation, calculated for a large enough temperature that quantum oscillations are washed out ($T = 1 \times \hbar v_F n_0^{1/3}/k_\textrm{B}$).  Different curves are labeled according to their value of the scattering time $\tau$, with $\tau_0 = (v_F n_0^{1/3})^{-1}$.   At large $B$, $\alpha_{xy}$ saturates to the value given by Eq.\ (\ref{Eq:alphaD2}), indicated by the dashed line.  In both plots, the units of magnetic field are $B_0 = \hbar n_0^{2/3}/e$, and units of $\alpha_{xy}$ are $\alpha_0 = e k_\textrm{B} n_0^{1/3}/\hbar$.}
\label{fig:alphaxyD}
\end{figure}

The general expression for $\alpha_{xy}$ in the dissipationless limit is given by
\begin{multline}
\alpha_{xy}^D =  \frac{e N_f}{2\pi \hbar} \sum_{n=0}^{\infty}{}^{'} \int_0^{\infty} \frac{d k_z}{\pi} \left[ s\left( \frac{\ve_n^0(k_z) - \mu}{k_B T} \right) + \right. \\ \left. + s\left( \frac{\ve_n^0(k_z) + \mu}{k_B T} \right) \right], \label{Eq:alphaD}
\end{multline}
where the notation $\sum_{n=0}^{\infty}{}^{'}$ is used to mean that there is an extra factor $1/2$ multiplying the $n=0$ term of the sum, and $\ve_0^0(k_z)$ should be understood as $\ve_+^0(k_z)$ in the above expression. The first term inside the brackets of Eq.~(\ref{Eq:alphaD}) corresponds to the electron contribution, while the second term is due to holes. The behavior of $\alpha_{xy}^D$ as a function of magnetic field is shown in Fig.\ \ref{fig:alphaxyD}.

In the limit of sufficiently weak magnetic field that many Landau levels are occupied, $B \ll E_F^2/(\hbar e v_F^2)$, and of sufficiently low temperature that $k_B T \ll E_F=  \hbar v_F (6 \pi^2 n_0/N_f)^{1/3}$, the density of states is well approximated by its zero-field, zero-temperature value, $\nu(\ve) \approx N_f \ve^2/2\pi^2 \hbar^3 v_F^3$, and the chemical potential coincides with the Fermi energy at zero field, $\mu \approx E_F$. The thermoelectric Hall coefficient is then given by
\be
\alpha^D_{xy} \approx \left( \frac{N_f \pi^4}6 \right)^{1/3}\frac{k_B^2 T n_0^{2/3}}{\hbar v_F B}.
\label{Eq:alphaD1}
\ee

On the other hand, when the magnetic field is made strong enough that $B \gg E_F^2/(\hbar e v_F^2)$, the system enters the extreme quantum limit, in which only the zeroth Landau level contributes to $\alpha_{xy}$. In this limit the chemical potential is given by
\be
\mu \approx \frac{4\pi^2 \hbar^2 v_F n_0}{N_f B e},
\ee
leading to a thermoelectric Hall coefficient
\be
\alpha^{D}_{xy} \approx \frac{\pi^2}3 \frac{e k_B^2 T N_f}{(2\pi\hbar)^2 v_F}.
\label{Eq:alphaD2}
\ee
Strikingly, and unlike in the Schr\"{o}dinger case, at large magnetic fields $\alpha^D_{xy}$ does not decay to zero and it retains no dependence on the electron density or the Fermi energy.  Instead, $\alpha_{xy}$ plateaus at large magnetic field, with the quantity $\alpha_{xy} v_F/T$ achieving a quantized value that depends only on universal constants and on the number $N_f$ of fermion flavors.  (In cases of anisotropic Fermi velocity, the relevant value of $v_F$ in this expression is the velocity in the magnetic field direction.) This quantized result for $\alpha_{xy}$ is valid so long as the Landau level spacing $v_F \sqrt{\hbar e B}$ is much larger than both the zero-field Fermi energy $E_F$ and the thermal energy $k_B T$.

Finally, at large enough temperatures that $k_B T$ is larger than the Landau level spacing, $k_B T \gg v_F \sqrt{\hbar e B},\, E_F$, the chemical potential is given by 
\be
\mu \approx \frac{6 \hbar^3 v_F^3 n_0}{N_f k_B^2 T^2} \ll k_B T, \label{Eq:muhighT}
\ee
leading to a thermoelectric Hall coefficient
\be
\alpha_{xy}^D \approx \frac{7 \pi^2}{90} \frac{k_B^4 T^3 N_f}{\hbar^3 v_F^3 B}.  \label{Eq:alphaD3}
\ee
As in the low-temperature case, Eq.~(\ref{Eq:alphaD2}), there is no dependence on the electron concentration.

As in the case of Schr\"{o}dinger particles, the thermoelectric Hall coefficient varies as $\alpha_{xy}^D \propto 1/B$ at small fields in the dissipationless limit.  This divergence is truncated, however, at sufficiently small magnetic fields that the $\omega_c \tau < 1$.  To describe this regime, we, again, use the quasiclassical approach of Sec.~\ref{sec:QC}. Focusing on the low-temperature limit $k_B T \ll E_F$, and assuming that many Landau levels are filled, $v_F \sqrt{\hbar e B} \ll E_F$, we directly apply the Sommerfeld expansion~(\ref{Eq:Sommerfeld}) to extract $\alpha_{xy}^D$. The density of states in this regime is given by  $\nu(\ve) \approx N_f \ve^2/2\pi^2 \hbar^3 v_F^3$, while the Fermi velocity $v_F$ is an energy-independent constant. An important difference with the Schr\"{o}dinger case is that the cyclotron frequency for Dirac electrons depends on energy: $\omega_c(\ve) = e B v_F^2 / \ve$. Collecting everything together, we find
\be
\alpha_{xy}^{DQC} \approx \frac{N_f}{18} \frac{e^2 k_B^2 T v_F \tau^2 B}{\hbar^3} \frac{ 1 + 3 \omega_c^2(E_F) \tau^2}{\left( 1 + \omega_c^2(E_F) \tau^2 \right)^2},
\ee
where the  Fermi energy is given by $ E_F =  \hbar v_F (6 \pi^2 n_0/N_f)^{1/3}$. In the limit of weak scattering, $E_F \ll e B v_F^2 \tau$, we reproduce Eq.~(\ref{Eq:alphaD1}) obtained for the dissipationless limit.  On the other hand, when the magnetic field is weak enough that $\omega_c(E_F) \tau \ll 1$, we arrive at
\be 
\alpha_{xy}^{DQC} \approx \frac{N_f}{18} \frac{e^2 k_B^2 T v_F \tau^2 B}{\hbar^3}.
\label{eq:alphaxyDsmallB}
\ee

\begin{figure}[htb!]
\centering
\includegraphics[width=1.0 \columnwidth]{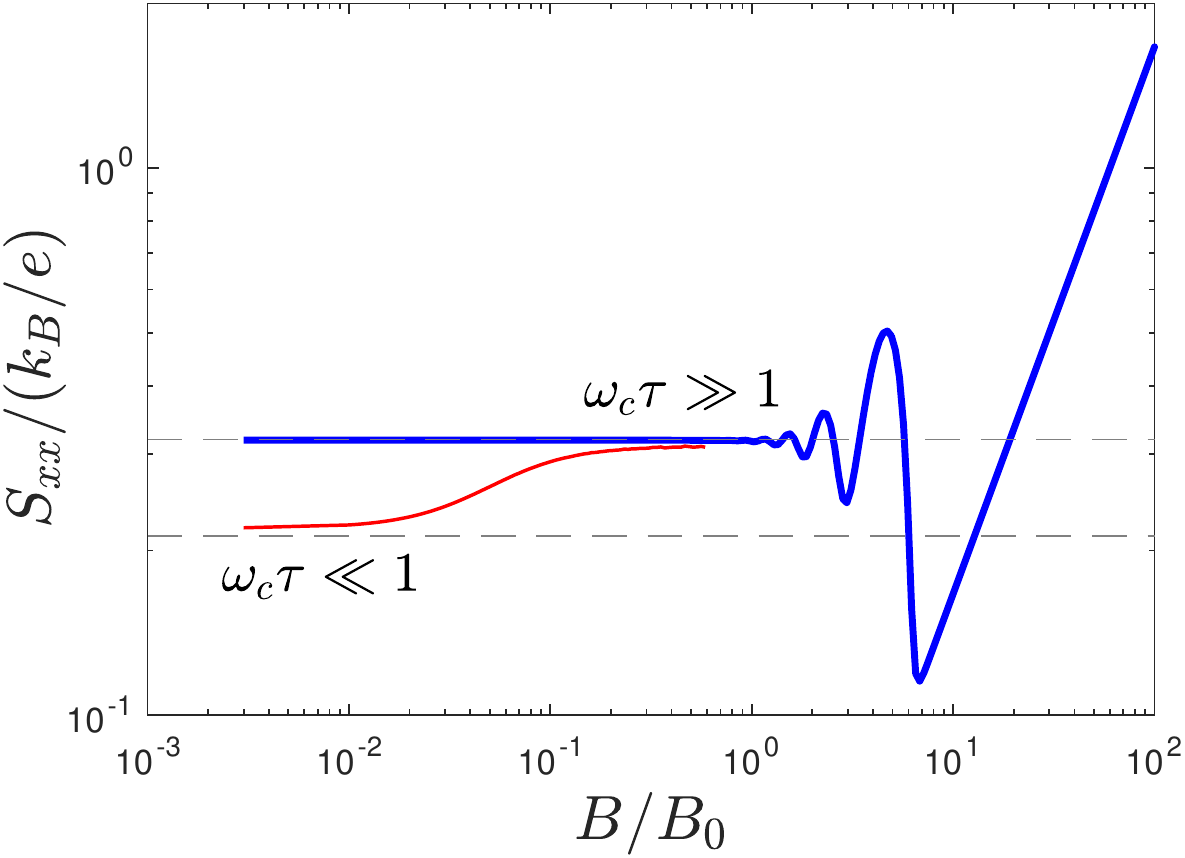}
\caption{The Seebeck coefficient $S_{xx}$ of a three-dimensional Dirac/Weyl semimetal as a function of magnetic field $B$.  The plot shows the result in the dissipationless limit (thick blue line) and the semiclassical result (thin red line) corresponding to a scattering time $\tau = 50 (v_F n_0^{1/3})^{-1}$.  The labeled dashed lines show the value of Eq.\ (\ref{eq:SxxDQC}) in the limits of $\omega_c \tau \ll 1$ and $\omega_c \tau \gg 1$, respectively.  The temperature is taken to be $T = 0.1 \hbar v_F n_0^{1/3}/k_\textrm{B}$. In both plots, the units of magnetic field are $B_0 = \hbar n_0^{2/3}/e$, and units of $\alpha_{xy}$ are $\alpha_0 = e k_\textrm{B} n_0^{1/3}/\hbar$. The linear increase in $S_{xx}$ with $B$ in the extreme quantum limit is described in detail in Ref.~\onlinecite{SkinnerFu}.}
\label{fig:SxxD}
\end{figure}

Let us now discuss the behavior of the Seebeck coefficient $S_{xx}$ and the thermodynamic figure of merit $ZT$ in Dirac materials. The behavior of $S_{xx}$ as a function of magnetic field is shown in Fig.~\ref{fig:SxxD}. As we show below, the energy-dependence of the cyclotron frequency in these materials has remarkable consequences for both $S_{xx}$ and $ZT$. Indeed, as is clear from Eq.~(\ref{Eq:SxxQC}), the quasiclassical expression for the Seebeck coefficient in the low-temperature and small-field limit is given by 
\be
S_{xx}^{DQC} \approx  \frac{\pi^2 k_B^2 T}{3 e E_F} \cdot \frac{2 + 3 \omega_c^2(E_F) \tau^2}{1 + \omega_c^2(E_F) \tau^2}.
\label{eq:SxxDQC}
\ee
(Here we have again assumed a constant scattering time~$\tau$.)
From this expression one can immediately see that the Seebeck coefficient at zero field is $3/2$ times smaller than that at $\omega_c(E_F) \tau \gg 1$.  Since the figure of merit $ZT$ of thermoelectric devices is proportional to $S_{xx}^2$ (see Ref.~\onlinecite{SkinnerFu} for a detailed discussion), a magnetic field for which $\omega_c \tau > 1$ produces a value of $ZT$ in Dirac materials that is enhanced by more than 100$\%$ relative to the zero-field case. Such a magnetic field is, in general, much weaker than the value required to achieve the extreme quantum limit, which was the primary focus of Ref.~\onlinecite{SkinnerFu}.  This enhancement of $ZT$ at relatively low fields should be contrasted with the case of Schr\"{o}dinger materials.  For such materials, as can be seen from Eq.~(\ref{Eq:SxxQC}), the Seebeck coefficient remains a constant at small fields and low temperatures $\hbar \omega_c(E_F), \, k_B T \ll E_F$, provided the scattering time is a constant.

We emphasize that the enhancement of $S_{xx}$ with magnetic field is a direct consequence of the dependence of the cyclotron frequency on energy. In case of an arbitrary (isotropic) dispersion relation $\ve(p)$, the solution of the Boltzmann equation gives a cyclotron frequency $\omega_c(\ve) = \left. e B [\ve'(p)/p]\right\vert_{p = p(\ve)}$. It is interesting to note, by examining Eq.~(\ref{Eq:SxxQC}), that in the case of a power-law dispersion $\ve(p) \propto p^\gamma$, $S_{xx}$ is enhanced by a weak ($\omega_c \tau \sim 1$) magnetic field if $\gamma<2$, and suppressed if $\gamma >2$. For Schr\"{o}dinger particles, $\gamma=2$, the Seebeck coefficient remains a constant at small magnetic fields.

Finally, in the limit of high temperature, $k_B T \gg v_F \sqrt{\hbar e B},\, E_F$, one cannot apply the Sommerfeld expansion~(\ref{Eq:Sommerfeld}) anymore, and one must use the general expression~(\ref{Eq:QC}) instead. Since chemical potential [Eq.\ (\ref{Eq:muhighT})] is small in this case, it can be neglected, leading to
\begin{align}
\alpha_{xy}^{DQC} &\approx \frac{N_f k_B^2 T v_F e^2 B \tau^2}{6 \pi^2 \hbar^3} \int_{-\infty}^{\infty}\frac{x^4 e^{x}}{(1+ e^{x})^2} \frac{dx}{x^2 + \omega_c^2(k_B T) \tau^2} \nonumber \\ &\approx \left\{\begin{array}{cc}  \frac{7 \pi^2}{90} \frac{k_B^4 T^3 N_f}{\hbar^3 v_F^3 B}, & \omega_c(k_B T)\tau \gg 1 \\ \frac{N_f k_B^2 T v_F e^2 B \tau^2}{18 \hbar^3}, & \omega_c(k_B T) \tau \ll 1  \end{array} . \right.
\end{align}
In the dissipationless limit, $\omega_c(k_B T) \tau \gg 1$, this expression agrees with the result of Eq.\ (\ref{Eq:alphaD3}).

\section{Summary and Discussion}
\label{sec:discussion}

In this paper we have presented a calculation of the thermoelectric reponse coefficients in Dirac and Weyl semimetals, focusing in particular on the thermoelectric Hall coefficient $\alpha_{xy}$ and the thermopower $S_{xx}$.  Our most notable results concern the enhancement of $\alpha_{xy}$ and $S_{xx}$ relative to the familiar case of Schr\"{o}dinger particles.  For example, applying a sufficiently strong field that $\omega_c \tau \gtrsim 1$ results in an enhancement of $S_{xx}$ [see Eq.\ (\ref{eq:SxxDQC})] that corresponds to a more than $100\%$ increase in the thermoelectric figure of merit $ZT$ (in a model where $\tau$ is energy-independent).  For Schr\"{o}dinger particles, on the other hand, there is no such enhancement.  At even larger fields, such that the chemical potential falls into the zeroth Landau level and the system enters the extreme quantum limit, $S_{xx}$ grows linearly with field without saturation.  This growth is accompanied by a striking plateau in $\alpha_{xy}$ [see Eq.\ (\ref{Eq:alphaD2})], such that the quantity $\alpha_{xy} v_F/T$ takes on a quantized value.  This is qualitatively different from the case of Schr\"{o}dinger particles, for which $\alpha_{xy}$ decays as $1/B$ at large fields and $S_{xx}$ saturates at a value of order $k_B/e$.

So far we are unaware of any published experimental measurements of $\alpha_{xy}$ in a Dirac or Weyl semimetal at large magnetic field.  However, the predictions of this paper should be readily testable in Dirac or Weyl semimetals with low electron density, such as Pb$_{1-x}$Sn$_x$Se \cite{OngPbSnSe} or ZrTe$_5$ \cite{OngZrTe5, ZhangZrTe5}. The enhancement of $S_{xx}$ with magnetic field was observed in Pb$_{1-x}$Sn$_x$Se in Ref.\ \onlinecite{OngPbSnSe}.

Throughout the work, we assumed that the main contribution to the thermoelectric coefficients is either from a single Dirac or Schr\"{o}dinger band. For many materials, however, there are multiple bands intersecting the Fermi level, and each of these provides a contribution to the thermoelectric response. Since the effects studied in our work are essentially single-particle phenomena, the contributions to $\alpha_{xy}$ and $S_{xx}$ from different bands simply add up, so the generalization to this case is straightforward.

A natural extension of the work presented here is to the case of a massive Dirac dispersion, for which the zero-field dispersion relation has the form $E_{\pm} = \pm \sqrt{(\Delta/2)^2 + \hbar^2 v_F^2 k^2}$.  (Here, the labels $\pm$ refer to the conduction and valence bands, respectively, and $\Delta$ is the energy gap between them.) While an exact calculation for this case is left for a later work, one can generally expect the thermoelectric behavior for such gapped Dirac materials to be similar to either the gapless Dirac case or the Schr\"{o}dinger case, depending on whether the thermal energy and the Fermi energy are large or small compared to $\Delta$.  In particular, if $\kb T \gg \Delta$, then the band gap is unimportant and one can describe both $\alpha_{xy}$ and $S_{xx}$ using the results in Sec.\ \ref{sec:Dirac}.  On the other hand, if both $\kb T$ and the zero-field Fermi energy $E_F$ are much smaller than $\Delta$, then the thermoelectric response is dominated by the low-momentum states near the band edge, for which the energy varies quadratically with momentum, and the thermoelectric response is well-described by the Schr\"{o}dinger-case results of Sec.\ \ref{sec:Sch}.  In the case where $E_F \gg \Delta \gg k_B T$, then at zero magnetic field the chemical potential is high in the conduction band, and the system behaves like a Dirac system (Sec.\ \ref{sec:Dirac}).  However, at high enough magnetic field that $B \gg \hbar^2 v_F n_0/(N_f e \Delta)$, the chemical potential falls and closely approaches the bottom of the conduction band, and the system behaves as in the Schr\"{o}dinger case (Sec.\ \ref{sec:Sch}). We expect that this crossover from Dirac-like to Schr\"{o}dinger-like behavior with increasing magnetic field can be relevant to Pb$_{1-x}$Sn$_x$Te and PbTe, where the band gap can reach 0.2-0.3 eV~\cite{Strauss66}.  It should be noted that the crossover from a ``massless'' to a ``massive'' Dirac case may naturally occur at sufficiently high magnetic fields in Weyl semimetals, which necessarily host multiple nodes. Indeed, when the inverse magnetic length $ l_B^{-1} = \sqrt{ eB/\hbar}$ becomes comparable to the separation between nodes in the momentum space,  the tunneling between the zeroth Landau levels associated with the Weyl points of different chirality may cause splitting and open up an energy gap~\cite{Patrick2017,Jia2017,Kim2017}. In this case, the Dirac mass $\Delta$ will strongly depend on the magnetic field $B$.

It is also worth commenting on the case of layered Dirac materials, which resemble a stack of two-dimensional Dirac systems with a weak interlayer coupling energy $t$.  In cases where $\kb T \gg t$, the interlayer coupling can be neglected and the system is accurately described as a stack of independent two-dimensional systems.  In this case one can describe the thermoelectric Hall conductivity by using the theory of Girvin and Jonson \cite{GirvinJonson} and dividing the value of $\alpha_{xy}$ for the two-dimensional case by the interlayer spacing.  Such a description may be relevant to recent experiments in ZrTe$_5$, where a three-dimensional quantum Hall effect was recently discovered \cite{ZhangZrTe5}, and to graphite, in which a large Nernst effect has been observed \cite{graphite}. 

Finally, let us comment in more detail on the dependence of our results on disorder.  In two-dimensional quantum Hall systems, the values of $\alpha_{xy}$ and $S_{xx}$ are affected by disorder, since the presence of disorder tends to broaden the Landau levels and therefore reduce the electron entropy when a given Landau level is partially filled \cite{GirvinJonson2, KunYang}.  In contrast, our results for $\alpha_{xy}$ and $S_{xx}$ in Dirac/Weyl semimetals at large magnetic field are essentially unaffected by disorder.  This independence of $\alpha_{xy}$ and $S_{xx}$ on disorder can be understood as a consequence of a density of states that has no dependence on energy in the high-field limit. Indeed, in the extreme quantum limit in a Dirac/Weyl semimetal, the density of states becomes an energy-independent constant, $\nu \sim 1/(\hbar v_F l_B^2)$.  Consequently no ``broadening'' of the Landau level by disorder can affect its value, provided that the Landau level spacing $\hbar v_F/l_B$ is much larger than the disorder energy scale $\hbar/\tau$.

\acknowledgments 

The authors thank Liyuan Zhang and Xiaosong Wu for collaboration on a related experimental study.
This work is supported by DOE Office of Basic Energy Sciences, Division of Materials Sciences and Engineering under Award DE-SC0018945. LF is partly supported by the David and Lucile Packard Foundation. BS is supported by the NSF STC ``Center for Integrated Quantum Materials" under Cooperative Agreement No.\ DMR-1231319.

\bibliographystyle{apsrev}

\end{document}